\documentclass[epsfig,12pt]{article}
\usepackage{graphicx}
\usepackage{epsfig}
\usepackage{graphicx}
\usepackage{amssymb}
\usepackage{amsmath}
\usepackage{mathtools}
\usepackage{hyperref}
\usepackage{tikz}
\usepackage[nottoc,notlof,notlot]{tocbibind} 
\usepackage[titles]{tocloft}

\usepackage{array}

\newcommand{\pdvp}{\partial_{++}}
\newcommand{\pdvm}{\partial_{--}}

\newcommand{\beq}{\begin{equation}}   
\newcommand{\eeq}{\end{equation}}
\newcommand{\beqn}{\begin{eqnarray}}   
\newcommand{\eeqn}{\end{eqnarray}}

\newcommand{\mtr}[3]{#1_{#2\bar{#3}}\,}
\newcommand{\imtr}[3]{#1^{#2\bar{#3}}\,}
\newcommand{\mcn}[3]{\Gamma^{#1}_{#2#3}}
\newcommand{\dlt}[2]{\delta^{#1}_{#2}}
\newcommand{\olra}[1]{\overset\leftrightarrow{#1}}
\newcommand{\riec}[4]{R_{\bar{#1}#2#3}^{~ ~ ~ #4}}

\begin{document}
\unitlength = 1mm

\def\de{\partial}
\def\Tr{ \hbox{\rm Tr}}
\def\const{\hbox {\rm const.}}  
\def\o{\over}
\def\im{\hbox{\rm Im}}
\def\re{\hbox{\rm Re}}
\def\bra{\langle}\def\ket{\rangle}
\def\Arg{\hbox {\rm Arg}}
\def\Re{\hbox {\rm Re}}
\def\Im{\hbox {\rm Im}}
\def\diag{\hbox{\rm diag}}


\def\Tr{{\rm Tr}}
\def\res{{\rm res}}
\def\Bf#1{\mbox{\boldmath $#1$}}
\def\balpha{{\Bf\alpha}}
\def\bbeta{{\Bf\beta}}
\def\bgamma{{\Bf\gamma}}
\def\bnu{{\Bf\nu}}
\def\bmu{{\Bf\mu}}
\def\bphi{{\Bf\phi}}
\def\bPhi{{\Bf\Phi}}
\def\bomega{{\Bf\omega}}
\def\blambda{{\Bf\lambda}}
\def\brho{{\Bf\rho}}
\def\bsigma{{\bfit\sigma}}
\def\bxi{{\Bf\xi}}
\def\bbeta{{\Bf\eta}}
\def\d{\partial}
\def\der#1#2{\frac{\d{#1}}{\d{#2}}}
\def\Im{{\rm Im}}
\def\Re{{\rm Re}}
\def\rank{{\rm rank}}
\def\diag{{\rm diag}}
\def\2{{1\over 2}}
\def\ntwo{${\mathcal N}=2\;$}
\def\nfour{${\mathcal N}=4\;$}
\def\none{${\mathcal N}=1\;$}
\def\ntwot{${\mathcal N}=(2,2)\;$}
\def\ntwoo{${\mathcal N}=(0,2)\;$}
\def\x{\stackrel{\otimes}{,}}

\newcommand{\cpn}{CP$(N-1)\;$}
\newcommand{\wcpn}{wCP${N,\widetilde{N}}(N_f-1)\;$}
\newcommand{\wcpd}{wCP$_{\widetilde{N},N}(N_f-1)\;$}
\newcommand{\vp}{\varphi}
\newcommand{\pt}{\partial}
\newcommand{\tN}{\widetilde{N}}
\newcommand{\ve}{\varepsilon}
\newcommand{\vr}{\varrho}
\renewcommand{\theequation}{\thesection.\arabic{equation}}

\newcommand{\sun}{SU$(N)\;$}

\setcounter{footnote}0

\vfill

\begin{titlepage}

\begin{flushright}
FTPI-MINN-19/15, UMN-TH-3824/19\\
\end{flushright}

\vspace{2mm}

\begin{center}
{  \Large \bf  
From Gauged  Linear Sigma Models to  Geo-\\[2mm] metric  Representation of  $\mathbb{WCP}\boldmath{(N,\tilde{N})}$ in 2D 
}

\vspace{5mm}

{\large \bf   Chao-Hsiang Sheu$^{a}$ and Mikhail Shifman$^{a, b}$}
\end {center}

\begin{center}

    {\it  $^{a}$Department of Physics,
University of Minnesota,
Minneapolis, MN 55455}\\{\small and}\\
{\it  $^{b}$William I. Fine Theoretical Physics Institute,
University of Minnesota,
Minneapolis, MN 55455}\\
\end{center}

\vspace{1cm}

\begin{center}
{\large\bf Abstract}
\end{center}
In this paper two issues are addressed. First, 
we discuss renormalization properties of a class of gauged linear sigma models (GLSM) which reduce to $\mathbb{WCP}\boldmath{(N,\tilde{N})}$ non-linear sigma models (NLSM)  in the low-energy limit. Sometimes they are referred to as the Hanany-Tong models. If supersymmetry is ${\cal N} =(2,2)$ the ultraviolet-divergent logarithm in LGSM appears, in the renormalization of the Fayet-Iliopoulos parameter, and is exhausted by a single tadpole graph. This is not the case in the daughter NLSMs. As a result, the one-loop renormalizations are different in GLSMs and their daughter NLSMs We explain this difference and identify its source.

In particular, we show why at $N=\tilde N$ there is no UV logarithms in the parent GLSM, while they do appear on the corresponding NLSM does not vanish. In the second part of the paper we discuss the same problem for a class of  ${\cal N} =(0,2)$ GLSMs considered previously. In this case renormalization is not limited to one loop;  all-orders exact  $\beta$ functions for GLSMs are known. We discuss divergent loops at one and two-loop levels.

\end{titlepage}

\newpage

\section{Introduction}
\label{intro}

In 1979 Witten suggested \cite{W1} an ultraviolet (UV) completion for $\mathbb{CP}(N-1)$, one of the most popular non-linear sigma models (NLSM),  with the aim  of large-$N$ solution of the latter. He considered both, non-supersymmetric and \ntwot versions. In the supersymmetric case the UV completion is in fact, a two-dimensional  scalar SQED with the Fayet-Iliopoulos term and judiciously chosen $n$ fields. UV completions of this type are referred to as gauged linear sigma models (GLSM). 

The target space of $\mathbb{CP}(N-1)$ and similar models (see below) is K\"ahlerian\,\footnote{More exactly, $\mathbb{CP}(N-1)$ is a particular case of the Grassmann model which, in turn,  belongs to the class of compact, homogeneous symmetric K\"ahler manifolds.} and of the {\em Einstein} type. Such models are renormalizable since all higher-order corrections are proportional to the target-space metric, and, therefore, are  characterized by a single coupling constant.\footnote{In (2,2) supersymmetric models the first loop is the only one which contributes to the coupling constant renormalization. In (0,2) models fermions do not contribute in the first loop,
manifesting themselves starting from two loops.} Thus, geometry of the target space is fixed up to a single scale factor.

The renormalization group (RG) flow from GLSM to NLSM is smooth, no change in the $\beta$ function occurs on the way.\footnote{Strictly speaking, whether this statement survives beyond one loop in models other than  ${\cal{N}}=(2,2)$ models is not fully known. This feature is by no means generic.} Moreover, for $\mathbb{CP}(N-1)$ we know the large-$N$ solution which explicitly matches the dynamical scale following from the $\beta$ function. For, say, the Grassmann model ${\mathcal G} (L,M)$ (here $M+L=N$) the solution is not worked out in full. However, the $\beta$ functions in both regimes -- GLSM and NLSM -- coincide \cite{ir,mor}.

In \cite{W2,V,HT} a generalization of  \ntwot GLSMs was suggested and discussed. These generalizations include a number of $n$ fields, with sign-alternating charges.
Of special importance is the case in which the number of positive charges $N$ is equal to that of the negative charges $\tilde N$.\footnote{The  general condition is
$\sum_i q_i +\sum_{\tilde i} \tilde q_{\tilde i} =0$.} In such GLSMs the Fayet-Iliopoulos
parameter is not renormalized (assuming ${\cal N}=(2,2)$). When these GLSMs are rewritten at low energies in the form of NLSMs they give rise to the so-called weighted  $\mathbb{WCP}(N,\tilde N)$ models. The target spaces in these cases are non-Einsteinian noncompact manifolds. Hence, these models are not renormalizable in the conventional sense of this word.\footnote{In \cite{ket} the notion
 of a generalized renormalizability of any two-dimensional  NLSM is presented in the form of a quantum deformation of its geometry described by the NLSM metric.
By nonrenormalizability we mean a more traditional definition that generally speaking an infinite number of counterterms is needed to eliminate all ultraviolet logarithms.
A thorough discussion of geometrical properties to be used below  can be found in \cite{kob}.}   We will discuss these \ntwoo models as well. Unlike \ntwot case in the $(0,2)$ models the second loop does not vanish in generic cases, resulting in ``new" structures. (In those special cases when it does, the third and higher loops do not vansis.) In this paper we address the issue of RG running in the parent-daughter pairs GLSM/NLSM for such target spaces. Discussion of some previous results in the Hanany-Tong model \cite{HT} which inspired the current work can be found in \cite{KS}. Recently, a number of GLSMs with sign-alternating charges were considered, and the $\beta$ functions calculated for (0,2) supersymmetric versions \cite{chens}.

Our conclusions are as follows.
For the class of GLSMs which upon reduction produce NLSMs of the $\mathbb{WCP}(N,\tilde N)$ type the RG evolution is more complicated and is not smooth. The K\"ahler potential (and, hence, the Lagrangian) of the resulting NLSMs
consists of two parts. The first part has the exactly the same structure as the second term in the bare K\"ahler potential $K^{(0)}$, see Eq. (\ref{eq:kp}). Its RG evolution produces the same formula for the renormalized coupling constant $r$  as for the FI constant in the parent GLSM. The first term in (\ref{eq:kp}) is not renormalized at all. Moreover, a new structure emerges upon RG evolution (see the  second line in Eq. (\ref{1kp})) which receives a logarithmic in $\mu$ coefficient in the RG flow, totally unrelated to that of $r(\mu)$. Thus,  the RG flow for the $\mathbb{WCP}(N,\tilde N)$ models is not described by a single running coupling constant. The number of the emergent structures will grow in higher loops in the {\em non-supersymmetric} case (see (\ref{A5})), so that these NLSMs are not renormalizable in the conventional sense of this word. Is the number of the emergent structures is limited in ${\cal N}=(0,2)$ supersymmetry? The answer
to this question can be found in Sec. \ref{seclo}.

The paper is organized as follows. In Sec. \ref{gencon}
 we briefly outline the GLSM formalism and renormalization of the FI constant under the RG evolution. Section \ref{geoboson} is devoted to reduction to NLSMs of 
 the $\mathbb{WCP}(N,\tilde N)$ type. We derive geometry of the target space: metric, Riemann and Ricci tensors, scalar curvature, etc. In Sec. \ref{sec4}
 we consider RG evolution in the $\mathbb{WCP}(N,\tilde N)$ models. Distinct structures responsible for different effects are isolated and a general result is formulated.
 Section \ref{simples} presents the simplest example of $\mathbb{WCP}(1,1)$ for illustration. In Sec. \ref{swcp} we work out the ${\cal N}=(0,2)$ versions of the 
 $\mathbb{WCP}(N,\tilde N)$ models.

\section{General Construction}
\label{gencon}

We start from presenting the bosonic part of our ``master'' model; its versions will be studied below. First, we introduce two types (or flavors) of complex fields $n_i$ and $\rho_a$, 
with the electric charges $+1$ and $-1$, respectively,
\beqn
S
& =&
\int d^2 x \left\{
 \left|\nabla_{\mu} n_{i}\right|^2 +  \left|\tilde\nabla_{\mu} \rho_{a}\right|^2  + \frac1{4e^2}F^2_{\mu\nu} + \frac1{e^2}
|\pt_\mu\sigma|^2+\frac1{2e^2}D^2
\right.
\nonumber\\[3mm]
 &+&    2|\sigma|^2\left(  |n_{i}|^2  + |\rho_a|^2\right) + iD \left(|n_{i}|^2 - |\rho_a|^2 -r\right)
\Big\} +\mbox{fermions}\,.
\label{cpg}
\eeqn
The index $i$ runs from $i=1,2,..., N$ while $a=1,2,...,\tilde{N}$. The action above is written in Euclidean conventions. 
The parameter $r$ in the last term of Eq. (\ref{cpg}) is dimensionless. It represents the two-dimensional Fayet-Iliopoulos term. 

The U(1) gauge field $A_\mu$ acts 
on $n$ and $\rho$ through appropriately defined covariant derivatives, \footnote{For generic situation, 
\begin{align}
	\nabla_{\mu}=\d_{\mu}-iq_{i}A_{\mu}\,, \qquad \widetilde{\nabla}_{\mu}=\d_{\mu}+i\tilde{q}_{a}A_{\mu},
\end{align}
which reduces to (\ref{22}) for $q_{i}=1$ and $\tilde{q}_{a}=-1$.
}
\beq\label{22}
 \nabla_{\mu}=\d_{\mu}-iA_{\mu}\,, \qquad \widetilde{\nabla}_{\mu}=\d_{\mu}+iA_{\mu}\,,
 \eeq
reflecting the sign difference between the charges. The electric coupling constant $e^2$ has dimension of mass squared. A key physical scale is defined through the
product
\beq
m^2_V = e^2 r\,.
\label{23t}
\eeq
If $e^2 \to \infty$ all auxiliary fields (i.e. 
$D$ and $\sigma$) 
can be integrated out, and we are in the NLSM regime. All terms except the kinetic terms of $n$ and $\rho$ disappear from the action, while the last term
reduces to the constraint
\beq
\sum_{i=1}^N |n_{i}|^2 - \sum_{a=1}^{\tilde{N}} |\rho_a|^2 =r\,.
\label{24}
\eeq

\begin{figure}[!t]
\centerline{\includegraphics[width=7cm]{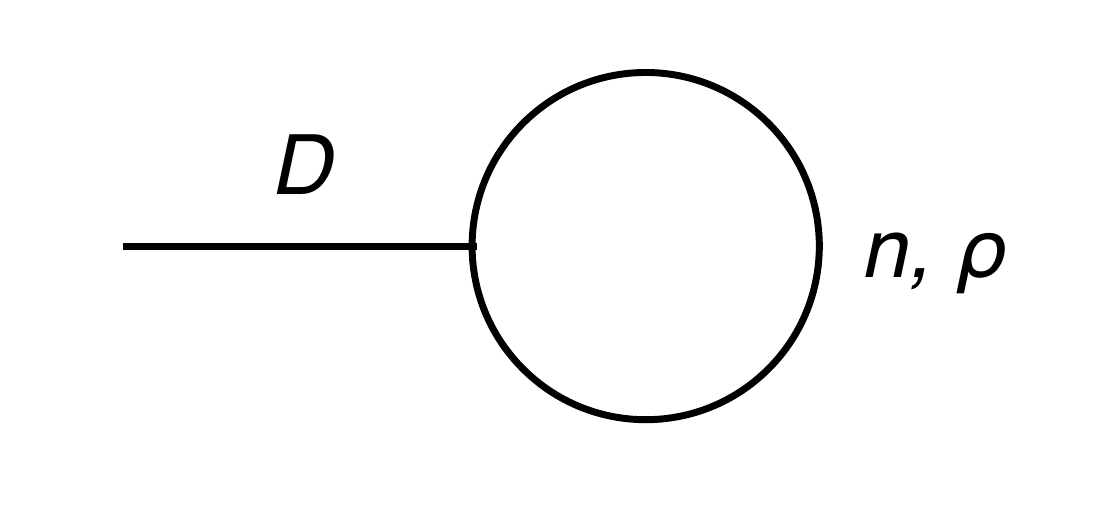}}
\caption{\small Tadpole graph determining renormalization of $r$ in the GLSM regime.}
\label{fig} 
\end{figure}

However, if the normalization point $\mu^2 \gg m^2$, the appropriate regime is that of GLSM.  The parameter $r$ is the only one which is logarithmically renormalized
at one loop in GLSM.  The only trivially calculable contribution comes from the tadpole
diagram of Fig. \ref{fig}. Namely,\footnote{Equation (\ref{25}) assumes that $r$ is positive and $N \geq \tilde{N}$. If $\tilde{N}>N$ 
one should consider negative $r$.} 
\beq
r(\mu ) = r_{\rm UV} - \frac{N-\tilde{N}}{2\pi}\log\frac{M_{\rm UV}}{\mu}\,.
\label{25}
\eeq
This renormalization vanishes if $N=\tilde N$ due to cancellation of charge $+1$ and $-1$ fields. Now we proceed to the discussion of the NLSM regime.

\section{Geometric formulation of $\mathbb{WCP}\boldmath{(N,\tilde{N})}$ }
\label{geoboson}
\setcounter{equation}{0}

To derive the geometric formulation we must take into account that the constrain (\ref{24}) and the U(1) gauge invariance
reduce the number of complex fields from $N+\tilde N$ in the set $\{n_i\}+ \{\rho_a\}$  down to $N+\tilde N -1$.
The choice of the coordinates on the target space manifolds can be made through various patches. For the time being we will choose one specific patch. Namely,
the last $\rho$ in the set $\{\rho_a\}$ (assuming it does not vanish on the selected patch) will be denoted as
\beq\label{36}
\rho_{\tilde N} =\varphi\,,
\eeq
where $\varphi$ will be set {\em real}. Then the coordinates on the target manifold are
\beqn
z_i &=& \varphi \, n_i ,\quad i=1,2,..., N\,, \nonumber\\[2mm] w_a &=& \frac{\rho_a}{\varphi},\quad a=1,2,...,\tilde{N}-1\,.
\label{35}
\eeqn
Note that the variables introduced in (\ref{35}) are not charged under U(1).

On the given patch $\varphi$ can be expressed in terms of the above coordinates as
\beq
\varphi =\left[ \frac{-r+H }{2\left(w_{a}\bar{w}_{a}+1\right)}  \right]^{1/2}
\label{38}
\eeq
where
\beq
H\equiv\sqrt{r^2+4z^{i}\bar{z}_{i}(w_{a}\bar{w}_{a}+1)}\,.  
\label{39}
\eeq
{\rule{0mm}{4mm}}The shorthand in Eq. (\ref{39}) will be used throughout the paper.
Integrating out the gauge field we observe that
\beq
	A_{\mu} = \frac{i}{2 H}
	\left( \frac{1}{\varphi^2}\,z_{i}\stackrel{\leftrightarrow}{\partial_\mu} {\bar{z}_{i}}
	-\varphi^2 \,w_{a}\stackrel{\leftrightarrow}{\partial_\mu} {\bar{w}_{a}}\right).
	\label{310}
	\eeq

Now we are ready to present the geometric data of the target space for $\mathbb{WCP}\boldmath{(N,\tilde{N})}$ in the following form:
\begin{align}\label{eq:wcpnmtr}
	\mtr{g}{i}{\jmath} = \frac{1}{\varphi^2}\left(\mtr{\delta}{i}{j}-\frac{1}{H\varphi^2}\bar{z}_{i}z_{\bar{\jmath}}\right), \quad 	\mtr{g}{i}{a} = \frac{1}{H} \bar{z}_{i} w_{\bar{a}}\, , \quad  \mtr{g}{a}{b} = \varphi^2 \left(\mtr{\delta}{a}{b}-\frac{\varphi^2}{H}\bar{w}_{a}w_{\bar{b}}\right),
	\end{align}
	with the inverse
	\begin{align}\label{ivmtr}
	\imtr{g}{i}{\jmath} = \varphi^2 \left(\delta^{i \bar{\jmath}} + \frac{1}{\varphi^4} \bar{z}^{\bar{\jmath}} z^{i}\right), \quad \imtr{g}{i}{a} = -\frac{1}{\varphi^2} z^{i} \bar{w}^{\bar{a}}\,, \quad	\imtr{g}{a}{b} = \frac{1}{\varphi^2} \left(\delta^{a \bar{b}} + w^{a} \bar{w}^{\bar{b}}\right).
	\end{align}
	This result can also be recovered by differentiating the corresponding K\"ahler potential \footnote{Equation (\ref{eq:kp}) coincides with (6.14) in Ref. \cite{SVY} if we take into account that with our coordinate patch $4\pi/g^{2}$ in \cite{SVY} should be replaced by $-r$.}
	\begin{align}\label{eq:kp}
	K^{(0)}(\phi_{p},\bar{\phi}_{\bar{q}}) = H + 2r\log{\varphi}
	\end{align}
	 which was previously found in \cite{KS,SVY}. We will consider the K\"ahler potential (\ref{eq:kp}) and its one-loop correction in more detail in the next section. For the time being, let us move on to further discuss geometry of this target space. To this end, we first find the metric connections,
	\begin{align} \notag
	\mcn{i}{j}{k} &= \frac{-1}{H\varphi^2}\left(\bar{z}_{j}\dlt{i}{k}+\bar{z}_{k}\dlt{i}{j}\right)+\frac{2}{H^2\varphi^4}z^{i}\bar{z}_j\bar{z}_{k}\,,\\[1mm] \notag
	\mcn{i}{a}{j} &= \frac{\varphi^2}{H} \bar{w}_{a} \dlt{i}{j}-\frac{2}{H^2}z^{i}\bar{w}_a\bar{z}_j \,,\\[1mm] \notag
	\mcn{i}{a}{b} &= \frac{2\varphi^4}{H^2}z^{i}\bar{w}_{a}\bar{w}_{b}\,,\\[1mm] \notag
	\mcn{a}{b}{c} &= -\frac{\varphi^2}{H}\left(\bar{w}_{b}\dlt{a}{c}+\bar{w}_{c}\dlt{a}{b}\right),\\[1mm]\notag
	\mcn{a}{b}{i} &= \frac{1}{H\varphi^2}\bar{z}_{i}\dlt{a}{b}\,,\\[1mm]
	\mcn{a}{i}{j} &= 0\,,\label{eq:mcnwpn}
	\end{align}
	where $ 1 \leq i, j \leq N$ and $1 \leq a, b \leq \tilde{N}-1$. Then  the Ricci tensor takes the form
	\begin{align}
	\mtr{R}{i}{\jmath} &= \frac{N-\tilde{N}}{r} \mtr{g}{i}{\jmath} \nonumber\\
	&+ \frac{(-r+H)\left[(\tilde{N}-N)H + r \right]}{rH^2\varphi^2 }\,\mtr{\delta}{i}{\jmath} - \frac{(-r+H)^2\left[(\tilde{N}-N)H+2r\right]}{rH^4 \varphi^4}\bar{z}_{i}z_{\bar{\jmath}}\,, \nonumber\\[2mm] 
	\mtr{R}{i}{a} &= \frac{N-\tilde{N}}{r} \mtr{g}{i}{a} + \frac1{rH^4}\left[ (\tilde{N}-N)H^3 +(\tilde{N}-N)Hr^2 +2r^3 \right] \bar{z}_{i} w_{\bar{a}}\,, \nonumber\\[2mm] 
	\mtr{R}{a}{b} &= \frac{N-\tilde{N}}{r}\,  \mtr{g}{a}{b}  \nonumber \\[2mm]
	& + \frac{\varphi^2(r+H)\left[(\tilde{N}-N)H + r \right]}{rH^2 }\,\mtr{\delta}{a}{b}- \frac{\varphi^4(r+H)^2\left[(\tilde{N}-N)H+2r\right]}{rH^4 }\bar{w}_{a} w_{\bar{b}} \,.
	\label{eq:wcpnric}
	\end{align}
	According to (\ref{eq:wcpnric}), the target space is \emph{not} an Einstein space. Furthermore, the scalar curvature is 
	\begin{align}\label{eq:scalar}
	R = \frac{2}{H^3}\left\{\left[(\tilde{N}-N)^2+(\tilde{N}+N-2)\right]H^2+2(\tilde{N} -N)H  r+2r^2\right\}.
	\end{align}
	Equation (\ref{eq:scalar}) implies that $H$ is a function of scalar curvature and parameters $N,\, \tilde{N}$ and $r$, say $H = H(R, r, N, \tilde{N})$. Note that setting $N$ equal to zero and $r$ negative,  we should be able to recover all well-known results in $\mathbb{CP}(\tilde{N}-1)$ model. Indeed, all non-Einstein terms in Eq. (\ref{eq:wcpnric}), the last line, vanish because $r+H=0$ (and so do the non-Einstein terms in the second and third lines, because we must put all term with $z_{i}$ to zero in (\ref{eq:wcpnric})). Then the coefficients of the Ricci and scalar curvature also match, namely,
\begin{align}
    \mtr{R}{a}{b} \to -\frac{\tilde{N}}{r}\mtr{g}{a}{b} \qquad {\rm and} \qquad R \to -\frac{2}{r}\tilde{N}\left( \tilde{N}-1 \right) \,,
\end{align}
with $r<0$.
	
Next, we observe that the general theory of NLSMs implies at one loop (see e.g. \cite{ket})
\beq
S_{\rm NLSM} = \int d^2 x\,  \left\{g_{p\bar{q}}\left( \partial_\mu \phi^p \partial_\mu \bar\phi^{\bar{q}}\right) - \frac1{2\pi}\log{\frac{M_{UV}}{\mu}} \, R_{p\bar{q}}\left( \partial_\mu \phi^p \partial_\mu \bar\phi^{\bar{q}}\right)\right\}
+...
\label{317}
\eeq	
where $\left\{\phi^{\,p} \right\}$ is generic coordinate of the target space. The question we will address now is the relation between two results:  Eq. (\ref{25}) in GLSM and Eq. (\ref{eq:wcpnric}) in NLSM. Both expressions mentioned above are known in the literature, (for (\ref{eq:wcpnric}) with a particular choice of $N,\,\tilde N$ see e.g. \cite{KS}). Clarification of their relationships is our starting goal.

\section{Renormalization in GLSM vs. NLSM}	
\label{sec4}
\setcounter{equation}{0}

In this section, we will study the renormalization of  $\mathbb{WCP}\boldmath{(N, \tilde{N})}$ model in the NLSM regime, and trace its origin from the parent GLSM. First of all, to discuss the renormalization structure, it is convenient to rephrase the previous results in terms of the K\"ahler potential.

For a K\"ahler manifold endowed with the K\"ahler potential $K(\phi_{p},\bar{\phi}_{\bar{q}})$, the metric is determined by the relation
\begin{align}\label{417}
	\mtr{g}{p}{q} = \frac\d{\d \phi^{p}}\frac\d{\d \bar{\phi}^{\bar{q}}}K(\phi_{p},\bar{\phi}_{\bar{q}}),
\end{align}
while other components vanish. The corresponding Ricci tensor is given by
\begin{align}\label{418}
	\mtr{R}{p}{q} = -\frac\d{\d \phi^{p}}\frac\d{\d \bar{\phi}^{\bar{q}}} \log{\sqrt{g}}
\end{align}
where $g$ represents the determinant of the metric tensor,
\begin{align}
	g = \left| \det{\left\{ \mtr{g}{p}{q}\right\}}\right|.
\end{align} If this manifold admits an Einstein-K\"ahler metric, the Ricci tenosr is propotional to its metric, in other words,
\begin{align}
	-\log{\sqrt{g}} = \alpha K(\phi_{p},\bar{\phi}_{\bar{q}})
\end{align}
for some constant $\alpha$. Yet, our case does \emph{not} belong to this class.

Back to our model, we can recover the result (\ref{eq:wcpnmtr}) by using (\ref{eq:kp}) and (\ref{417}). For convenience we will represent it in a different form,
\begin{align}\notag
	\mtr{g}{i}{\jmath} &=\frac{-r+H}{H\varphi^2}\mtr{\delta}{i}{\jmath} - \frac{(-r+H)^2}{H^3\varphi^4}\bar{z}_{i} z_{j} +r\underbrace{\left( \frac{1}{H\varphi^2}\mtr{\delta}{i}{\jmath} -\frac{-r+2H}{H^3 \varphi^4}\bar{z}_{i} z_{j} \right)}, \\[2mm]
	\mtr{g}{i}{a} &= \frac{H^2+r^2}{H^3}\bar{z}_{i}w_{\bar{a}} + \underbrace{\frac{-r^2}{H^3}\bar{z}_{i}w_{\bar{a}} },\\[2mm]
	\mtr{g}{a}{b} &= \frac{\varphi^2(r+H)}{H}\mtr{\delta}{a}{b} - \frac{\varphi^4(H+r)^2}{H^3}w_{\bar{b}}\bar{w}_{a} +\underbrace{ r\left(-\frac{\varphi^2}{H}\mtr{\delta}{a}{b} + \frac{\varphi^4(2H+r)}{H^3}w_{\bar{b}}\bar{w}_{a}\right) }\notag 
\end{align}
In the above formulas for the metric tensor (they are identical to (\ref{eq:wcpnmtr}))  we separate the contributions from $H$ and $2r\log\varphi$, respectively, in the K\"ahler potential $K^{(0)}$,
namely, the terms marked by underbrace  originate from $2r\log\varphi$ in Eq. (\ref{eq:kp}).

From the expression (\ref{eq:wcpnmtr}), the metric determinant can be calculated in a straightforward manner, and we obtain
\begin{align}\label{eq:md}
		-\log{\sqrt{g}} = 2(N-\tilde{N}) \log{\varphi} + \log{H} 
\end{align}
for which the result coincides with the example in \cite{KS} with a particular pair of $N, \, \tilde{N}$. As a consistency check, we can apply (\ref{418}) to (\ref{eq:md}) to see that it indeed reproduces (\ref{eq:wcpnric}). Also, it is instructive to explicitly indicate the Einstein part and non-Einstein part in the Ricci curvature. Namely,
\begin{align}\label{422}
	\mtr{R}{p}{q} = \left(\frac{N-\tilde{N}}{r}\right)\mtr{g}{p}{q} + \frac\d{\d \phi^{p}}\frac\d{\d \bar{\phi}^{\bar{q}}} \left(\log{H}-\frac{N-\tilde{N}}{r}H\right) .
\end{align}

At one-loop level, the K\"ahler potential acquires a correction following from (\ref{eq:md}), see also (\ref{422}),
\begin{eqnarray}\label{1kp}
K^{(0)}  + K^{(1)} &=& K^{(0)} -\frac{1}{2\pi}\log{\frac{M_{\rm UV}}{\mu}}\left[\frac{N-\tilde N}{r} K^{(0)}+\left(
\log H - \frac{N-\tilde N}{r}  H\right)\right]
\nonumber\\[2mm]
&=&
H+ 2\left( r- \frac{N - \tilde{N}}{2\pi}\log{\frac{M_{\rm UV}}{\mu}} \right)\log{\varphi} 
\nonumber\\[2mm]
&-& \left(\frac1{2\pi}\log{\frac{M_{\rm UV}}{\mu}}\right)\log{H} \,.
\end{eqnarray}
The correction of the coupling constant, $r$, and the $\log{H}$ term result from the first and the second terms in Eq. (\ref{eq:md}), respectively while the corrections to $H$ term cancel. As a consistency check, considering the $\mathbb{CP}(\tilde{N}-1)$ case ($N=0$), we observe that $H$ reduces to a constant and and dropping all $H$ terms  we observe that the K\"ahler potential renormalizes multiplicatively. We recover the  conventional $\mathbb{CP}(\tilde{N}-1)$ result. 

We can immediately read off from Eq. (\ref{1kp}) that the FI parameter is renormalized as\,\footnote{To factor out $r$ in $K^{(0)}$ it is convenient to rescale $z$, namely $z\to z r$. Then $H$ becomes proportional to $r$ and $K^{(0)}\to r \tilde{K}^{(0)}$ where   $\tilde{K}^{(0)}$ is $r$ independent.}
\begin{align}\label{eq:fr}
	r(\mu) = r_{\rm UV} - \frac{N - \tilde{N}}{2\pi}\log{\frac{M_{\rm UV}}{\mu}}\,,
\end{align}
in agreement with (\ref{25}) obtained in the GLSM analysis.

As an essential example, we consider the model with the equal numbers of positive and negative charges. Then, as shown in (\ref{eq:fr}), the FI parameter gets no correction and the corresponding $\beta$ function vanishes. However, the K\"ahler potential is still modified by the one-loop contribution,
\begin{align}
	\left. K^{(0)}  + K^{(1)} \right|_{N=\tilde{N}}= 2r\log{\varphi} + H - \left(\frac1{2\pi}\log{\frac{M_{\rm UV}}{\mu}}\right)\log{H}.
\end{align} 
The emergent term proportional to $\log H$ does not vanish even if $N=\tilde{N}$, therefore making the theory non-renormalizable (in the case of ${\cal N} = (0,2)$ supersymmetry, see Sec.~\ref{swcp}).

Note that in the generic case $N\neq  \tilde{N}$ but $N\sim \tilde{N}$ the renormalization of $r$ scales as $N$ while the coefficient of $\log H$ is $O(N^0)$.
Then the latter can be ignored in the large-$N$ limit.

\section{Where does the discrepancy between GLSM and NLSM come from?}

\begin{figure}[!t]
\centerline{\includegraphics[width=7cm]{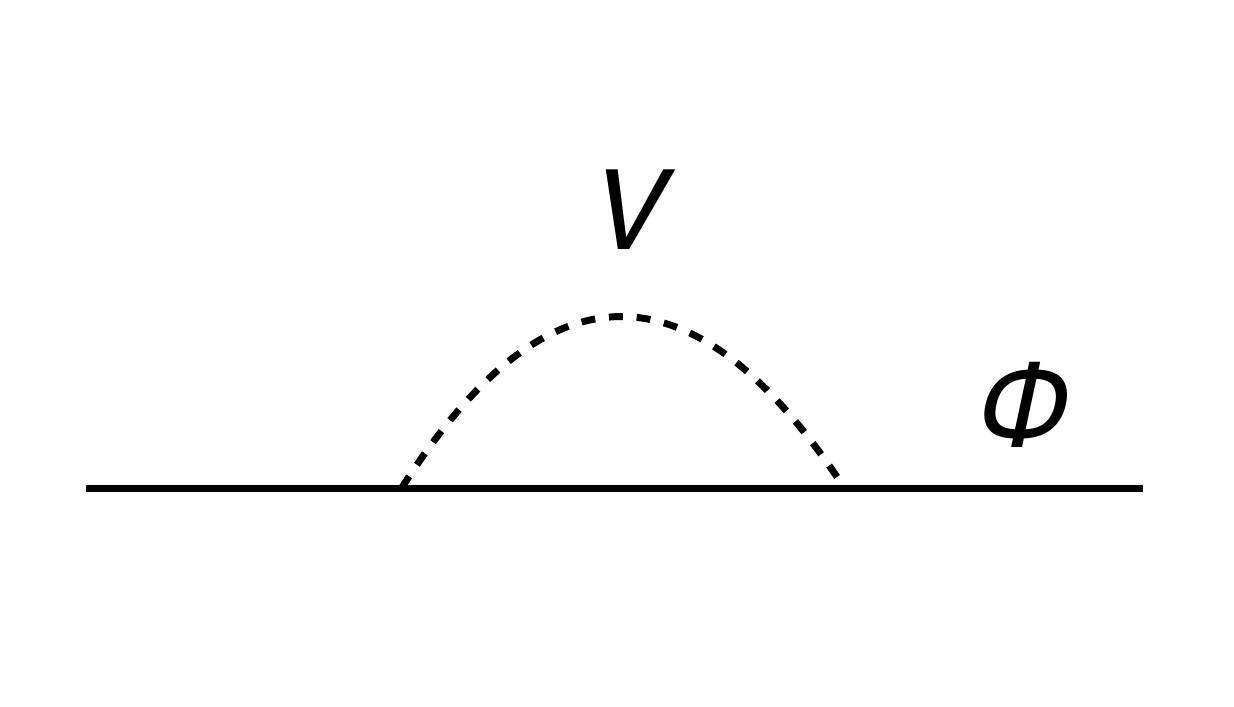}}
\caption{\small $Z$ factor of the matter fields in GLSM with $\mu \ll m_V$.}
\label{ququ} 
\end{figure}

The answer to the above question might seem paradoxical. Let us return to Sec. \ref{cpg} in which it was stated that
the only ultraviolet logarithm in GLSM comes from the Fayet-Iliopoulos term renormalization depicted in Fig. \ref{fig}. This statement is correct. However, this does not mean that there are no other logarithms in this model (with is a two-dimensional reduction of SQED with matter fields possessing  opposite charges). If we descend down in $\mu$ below $ m_V$ (see (\ref{23t})), we will discover logarithms of $m_V/\mu$ rather than $\log M_{\rm UV}/\mu$. The former in a sense might be called ``infrared." They come from the $Z $ factors of the matter fields in (\ref{cpg}) and are determined by the graphs shown in Fig. \ref{ququ}.

On dimensional grounds the one-loop contribution to the $Z$ factor is proportional to
\beq
e^2 m_V^{-2} \,\log \frac{m_V}{\mu} \sim \frac{1}{r}  \log \frac{m_V}{\mu}\,.
\eeq
It is curious that the same type of ``infrared" logarithms were found 45 years ago \cite{peng} in weak flavor-changing decays and are widely known now as penguins.
They are typical of theories with multiple scales.

In passing from GLSMs to NLSMs we tend $m_V\to \infty$ thus identifying it with $M_{\rm UV}$. The distinction between two types of logarithms are lost.

We conclude that the Fayet-Iliopoulos  parameter is related -- in the NLSM formulation -- to the cohomology class of the K\"ahler form of the target space generically defined as 
\begin{align}
	\omega = \frac{i}{2}\, \mtr{g}{p}{q} d\phi^{p} \wedge d\overline{\phi}^{\,\overline{q}},
\end{align}
where $d$ is the de Rham operator.\footnote{ See \cite{W2,hkp} for more thorough discussions. We thank  A. Gorsky, S. Ketov, A. Losev and D. Tong for instructive correspondence on this issue.} Speaking in physical terms, the K\"ahler class can be viewed as a product of a complexified scale parameter $r$ and an analog of  the appropriately normalized topological (or $\theta$) term. The  latter takes integer values.

These remarks explains the structure of the first line in Eq. (\ref{1kp}), as well as the emergence of extra logarithms. That's why in GLSM we recover Eq. (\ref{eq:fr}),
inherited from GLSM, in addition to an ``extra" last term in the first line of Eq. (\ref{1kp}).

\section{The simplest example: $\mathbb{WCP}(1,1)$ model}
\label{simples}
\setcounter{equation}{0}

To further illustrate our analysis, let us have a closer look at the minimal example consisting of only two chiral fields, one with the positive unit charge and the other with the negative unit charge  (i.e. $N=\tilde{N}=1$). Also, the appropriate number of fermi superfields can be included, so we can consider either ${\cal N} = (2,2)$ or
${\cal N} = (0,2)$ theories. The bare K\"ahler potential in this problem is presented in \cite{msbook}, Sec. 52.

 Note that in the given simplest case
\beq
H = \sqrt{r^2 + 4 z\bar{z}}\,.
\label{t51}
\eeq

First of all, let us examine  the structure of the vacuum manifold in the corresponding GLSM,
\beq
	|n|^2 - |\rho|^2 = r\,.
	\label{w51}
\eeq
This space is simply a four-dimensional hyperboloid. Gauging out a $U(1)$ phase we arrive at a two-dimensional target space in $\mathbb{WCP}(1,1)$ (two real dimensions).

\begin{figure}[!htbp]
	\centering
	\includegraphics[height=5cm]{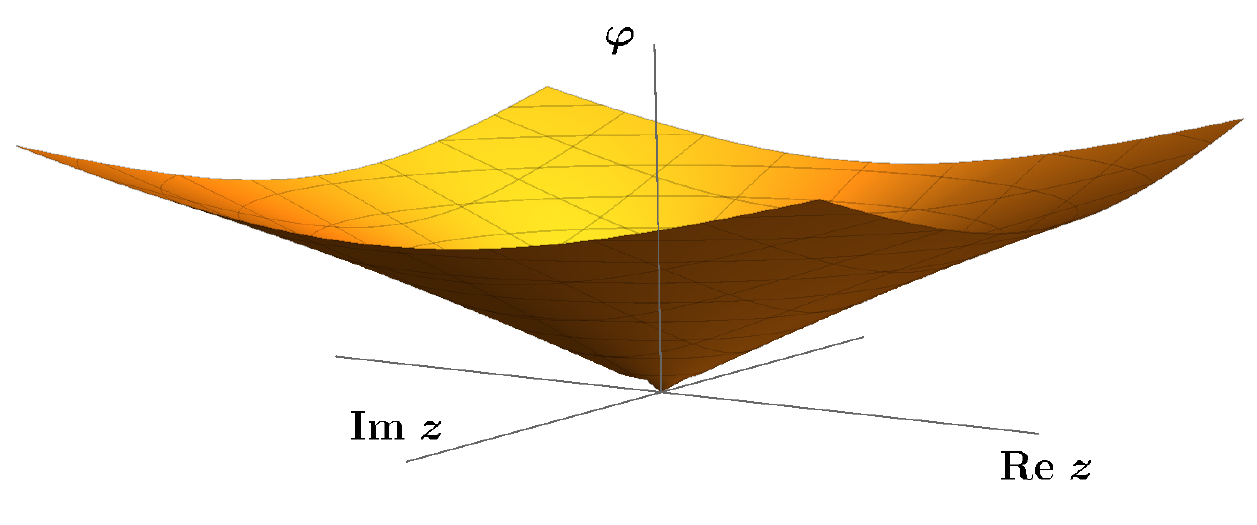}
	\caption{\small Geometry of the chosen patch of the target space in $\mathbb{WCP}(1,1)$ model.}
	\label{vm}
\end{figure}

Indeed, $\rho$ in Eq. (\ref{36}) can be chosen to be real and positive, then so is $\varphi$.  Using the choice of coordinates in (\ref{36}) and (\ref{35}), we can  reduce (\ref{w51})  to 
\begin{align}
	 \frac{| z |^2}{\varphi ^2 }- \varphi^2 = r,
	 \label{t53}
\end{align}
illustrated in Fig. \ref{vm}. From the graph, we see that the singularity at $\varphi = 0$ is one and the only one sigular point on this patch. However, considering $z$ and $\bar{z}$ as coordinates, we observe that 
\beq
\varphi^2 = -\frac{r}{2} + \sqrt {\frac{r^2}{4} +z\bar{z}}
\eeq
becomes zero at the origin, the point which must be punctured on the given patch.
 The constraints imposed on the fermion fields are of the type
\begin{align}
	\bar{n}\tau_{+} - \bar{\rho}\xi_{+}= 0\,,
\end{align}
 which implies that the fermions live on the tangent bundle of the target manifold, see Sec. \ref{swcp}.

Following the same line of calculation as in Sec. \ref{geoboson}, we obtain the only non-vanishing element of the metric 
\begin{align}\label{545}
	\mtr{g}{1}{1} = \frac1{\sqrt{r^2+4z\bar{z}}}\,.
\end{align}
Its connections are
\begin{align}
\Gamma^{1}_{11} &= -\frac{2\bar{z}}{r^2+4z\bar{z}},\quad \mbox{and} \quad \Gamma^{\bar{1}}_{\bar{1}\bar{1}} = -\frac{2z}{r^2+4z\bar{z}}\,.
\end{align}
In addition, it is not difficult to the curvature tensor,
\begin{align}\label{58}
R_{1\bar{1}1\bar{1}} = -\frac{2r^{2}}{\left(r^{2}+4z\bar{z}\right)^{5/2}}
\end{align}
and  the Ricci tensor,
\begin{align}\label{548}
R_{1\bar{1}} = \frac{2r^{2}}{\left(r^{2}+4z\bar{z}\right)^2}\, .
\end{align}
From (\ref{545}) and (\ref{548}), we can explicitly see that the Ricci tensor is not  proportional to the metric, and this is consistent with the fact that the target manifold is not of the Einstein type and our general analysis of Sec. \ref{sec4}. The scalar curvature is also computed, it reduces to
\begin{align}
R =  \frac{4r^{2}}{\left(r^{2}+4z\bar{z}\right)^{3/2}}\,.
\end{align}

Now, it is time to talk about the quantum correction of this model. That is, the $\beta$ function is computed as follows.
\begin{align}
	\beta(\mtr{g}{1}{1})_{\rm one-loop}= \frac{r^{2}}{\pi\left(r^{2}+4z\bar{z}\right)^2}.
\end{align}
For $\mathcal{N}=(2,2)$ case, this is the end of story. However, for a non supersymmetric model, it is not the case, i.e. it still receives the two-loop correction. Namely,
\begin{equation}
	\beta(\mtr{g}{1}{1})_{\rm two-loop}= 
	\frac{r^{2}}{\pi\left(r^{2}+4z\bar{z}\right)^2} \left[ 1 
	+ \frac{r^2}{2\pi\left(r^{2}+4z\bar{z}\right)^{3/2}} \right],
	\label{612t}
\end{equation}
see Sec. \ref{app} for various \ntwoo models. In this simplest example the Lagrangian (including one loop) can be written as follows:
\beq
{\cal L} = {\cal L}^{(0)}+ {\cal L}^{(1)}= \partial_\mu z\partial^\mu \bar{z}\, \left[ \frac{1}{H} -\left(\frac{r^2}{\pi} \log \frac{M_{\rm UV}}{\mu}\right) \, \frac{1}{H^4}\right],
\eeq
where
$H$ is given in (\ref{t51}).
The first term on the right-hand side represents the bare Lagrangian which is not renormalized (remember that in the case at hand $N=\tilde{N}=1$).  The second term emerges at  one loop -- a different structure proportional to $\log\mu$ which is absent in the UV. It can be ignored at large $|z|$.

As an aside, if we take the $U(1)$ charges of two chiral superfields to be $q$ and $-q$, the geometry of the target space does not change, but only the scale $r$ from the FI term rescales as $r/q$. For example,
\begin{align}
\mtr{g}{1}{1} = \frac{1}{\sqrt{(r/q)^2+4z\bar{z}}}\,.
\end{align}

\section{Generalization to \ntwoo $\mathbb{WCP}\boldmath{(N, \tilde{N})}$}\label{swcp}
\setcounter{equation}{0}

The \ntwoo deformation discussed in this section was introduced in \cite{chens}. With the fermion fields taken into account, we can work out  in this paper  heterotic supersymmetric versions. As suggested in \cite{Edt,tSY}, we construct \ntwoo GLSM with the gauge multiplet, fermi multiplets, and two types of the boson chiral superfields with $U(1)$ charge $+1$ and $-1$. 

Before proceeding to the invariant action, we recall the superfield representation for each multiplet. In this case, two chiral multiplets are
\begin{align}
	\mathcal{N}_{i} &= n_{i} + \sqrt{2} \theta^{+}\tau_{+,i}  -i \theta^{+}\bar{\theta}^{+}\nabla_{++} n_{i}, \nonumber\\[2mm]
	\varrho_{a} &= \rho_{a} + \sqrt{2} \theta^{+}\xi_{+,a}  -i \theta^{+}\bar{\theta}^{+}\nabla_{++} \rho_{a},
\end{align}
the gauge multiplet is
\begin{align}
	\mathcal{U}_{--} =  \sigma - 2i \theta^{+}\bar{\lambda}_{i} -2i\bar{\theta}^{+}\lambda_{i}  + 2 \theta^{+}\bar{\theta}^{+}D,
\end{align}
and, lastly, the fermi multiplets are
\begin{align}
	\Gamma_{-,M} = \chi_{-M} - \sqrt{2} \theta^{+}G_{M}  -i \theta^{+}\bar{\theta}^{+}\nabla_{++} \chi_{-M}.
\end{align}

Now, we are allowed to present the full expression of \ntwoo extension of the key model, which is non-minimal
\begin{align}\label{eq:hetwcpn}
	S_{(0,2)} &= \int d^2 x \left\{ 
	\left|\nabla_{\mu} n_{i}\right|^2 - i\bar{\tau}_{+,i}\nabla_{--}\tau_{+}^{i} +  \left|\tilde\nabla_{\mu} \rho_{a}\right|^2  - i\bar{\xi}_{+,a}\tilde{\nabla}_{--}\xi_{+}^{a} + \frac1{4e^2}F^2_{\mu\nu} \right.
	\nonumber\\[3mm]
	& + \frac1{e^2}~i\bar{\lambda}_{-}\nabla_{++}\lambda_{-} +\frac1{2e^2}D^2 - i \bar{\chi}_{-M}\nabla_{++}\chi^{M}_{-} + \left| G_M \right|^2 + \sqrt{2}\,\bar{n}_{i}\lambda_{-}\tau_{+}^{i}    \notag\\[3mm]
	&- \sqrt{2}\,\bar{\tau}_{+,i}\bar{\lambda}_{-}n^{i} - \sqrt{2}\,\bar{\rho}_{a}\lambda_{-}\xi_{+}^{a} +  \sqrt{2}\,\bar{\xi}_{+,a}\bar{\lambda}_{-}\rho^{a} + iD \left(|n_{i}|^2 - |\rho_a|^2 -r\right)
	\Big\} .
\end{align}
The covariant derivative for $\chi_{-M}$ field is defined though its $U(1)$ charge, $q_M$, such that
\begin{align}
	\nabla_{++}\chi_{-M} =  \left( \pt_{++} - iq_M A_{++} \right) \chi_{-M}
\end{align}
Note that $\sigma$ field is suppressed preserving $(0,2)$ supereymmetry. Also, we notice that $G_M$ is an auxiliary field.

In NLSM regime, the gauge multiplet becomes auxiliary (all kinetic terms vanish in $e^2 \to \infty$ limit), so the corresponding component fields (i.e. $\sigma$, $\lambda_{-}$, and $D$) can be integrated out to give the constraints. To be more precise, $D$ term again results in Eq. (\ref{24}), and gauginos yield
\begin{align}\label{529}
	\sum_{i=1}^{N} \bar{n}_{i}\tau_{+}^{i} - \sum_{a=1}^{\tilde{N}} \bar{\rho}_{a}\xi_{+}^{a} = 0
\end{align}
where the same condition applies to its hermitian conjugate.

To obtain the geometric formulation of \ntwoo $\mathbb{WCP}(N,\tilde{N})$, we follow the parallel treatment in section \ref{geoboson} to eliminate $U(1)$ redundancy by setting
\begin{align}
	\varrho_{\tilde{N}} = \varphi + \sqrt{2}\theta^{+} \kappa_{+} + \cdots
\end{align}
in which $\varphi$ is a \emph{real} field and $\kappa_{+}$ is a \emph{complex} Weyl fermion. Notice that $\varrho_{\tilde{N}}$ is assumed to be nowhere vanishing on the chosen patch. On the target manifold, bosonic coordinates are defined in the same way as Eq. (\ref{35}) while the fermionic coordinates are
\begin{align}
	\zeta_{+,i} &= \kappa_{+} n_{i} + \tau_{+,i}\varphi &&{\rm for ~} i=1,2,..., N\, , \nonumber\\[2mm]
	\eta_{+,a} &= \frac{1}{\varphi} \left(  \xi_{+,a} - \frac{\rho_{a}}{\varphi}\kappa_{+} \right) &&{\rm for ~} a=1,2,..., \tilde{N}-1\,.
\end{align}
This can be seen by taking the following parametrization for superfields
\begin{align}
	Z_{i} &= z_{i} +  \sqrt { 2 } \theta^{+} \zeta_{+,i} =:  \mathcal{N}_{i} \, \varrho_{\tilde{N}} 
	\nonumber\\[2mm]
	W_{a} &= w_{a} +  \sqrt { 2 } \theta^{+} \eta_{+,a} =:  \varrho_{a} \, \varrho_{\tilde{N}}^{-1}. \quad 
\end{align}

On this patch $\varphi$ has the identical expression as Eq. (\ref{38}) and $\kappa_{+}$ is written in terms of the above coordinates by
\begin{align}
	 \kappa_{+} = \frac{\bar{z}_i\zeta_{+}^{i}-\varphi^{4}\bar{w}_a \eta_{+}^{a}}{H\varphi}.
\end{align} 
Integrating out gauge fields we then find that
\begin{align}\label{533}
	A_{--} &=\frac{i}{2H}~\left(\frac{1}{\varphi^2}z^{i}\olra{\pdvm}{\bar{z}_{i}}-\varphi^2 w^{a}\olra{\pdvm}{\bar{w}_{a}}\right) -\frac{1}{H}~\sum_{M} q_{M} \overline{ \chi }_{-M}\chi_{-}^{M}\,, \nonumber\\[2mm]
	A_{++} &=\frac{i}{2H}~\left(\frac{1}{\varphi^2}z^{i}\olra{\pdvp}{\bar{z}_{i}}-\varphi^2 w^{a}\olra{\pdvp}{\bar{w}_{a}}\right) \nonumber\\[2mm]
	&\qquad \qquad +\mtr{h}{i}{\jmath}\bar{\zeta}_{+}^{\bar{\jmath}}\zeta_{+}^{i}+\mtr{h}{a}{b}\bar{\eta}_{+}^{\bar{b}}\eta_{+}^{a}+ \left(\mtr{h}{a}{\imath} \bar{\zeta}_{+}^{i}\eta_{+}^{a} + \mbox{h.c.}\right),
\end{align}
where 
\begin{align}
	\mtr{h}{i}{\jmath} &= \frac{-1}{H\varphi^2}\left(\mtr{\delta}{i}{j}-\frac{2H-r}{H^2\varphi^2}\bar{z}_{i}z_{\bar{\jmath}}\right), \nonumber\\[2mm]
	\mtr{h}{a}{b} &= \frac{\varphi^2}{H} \left(\mtr{\delta}{a}{b}-\frac{\varphi^2}{H^2} (2H+r)\bar{w}_{\bar{a}}w_{\bar{b}}\right),  \nonumber\\[2mm]
	\mtr{h}{a}{\imath} &= \frac{r}{H^3} \, z_{\bar{\imath}}\, \bar{w}_{a}. 
\end{align}
As a remark, these coefficients can also be related to the connection associated with $\chi_{M}$ fields (see Eq. (\ref{chc}) with $\pt_{++}$ replaced by the exterior derivative $d$) in the way
\begin{align}
	d\Omega = -i \mtr{h}{p}{q}\, d\phi^{p} \wedge d\overline{\phi}^{\,\bar{q}}.
\end{align}

Next, we present the final expression of the geometric formulation of \ntwoo $\mathbb{WCP}(N,\tilde{N})$ by collecting above ingredients.
\begin{align}\label{eq:nlsm}
S_{\rm NLSM} &= \int d^2 x \left\{ \mtr{g}{i}{\jmath}\pt_\mu \bar{z}^{\bar{\jmath}} \, \pt^{\mu}z^{i} + \mtr{g}{a}{b}\pt_\mu \bar{w}^{\bar{b}} \, \pt^{\mu} w^{a}  + \mtr{g}{i}{a}\pt_\mu \bar{w}^{\bar{a}} \, \pt^{\mu}z^{i} +\mtr{g}{a}{\imath}\pt_\mu \bar{z}^{\bar{\imath}} \, \pt^{\mu}w^{a} 
\right.
\nonumber\\[3mm]
& + i \mtr{g}{i}{\jmath} \bar{\zeta}^{\bar{\jmath}}_{+} \, \nabla^{c}_{--}\zeta^{i}_{+} + i \mtr{g}{a}{b} \bar{\eta}_{+}^{\bar{b}} \, \nabla^{c}_{--} \eta_{+}^{a}  + i \mtr{g}{i}{a} \bar{\eta}^{\bar{a}}_{+} \, \nabla^{c}_{--} \zeta^{i}_{+} + i \mtr{g}{a}{\imath} \bar{\zeta}^{\bar{\imath}}_{+} \nabla^{c}_{--} \eta_{+}^{a} \nonumber\\[3mm]
&+ i \bar{\chi}_{-M}\nabla^{f}_{++}\chi^{M}_{-} \nonumber\\[3mm]
&+ \left[ \mtr{h}{i}{\jmath} \bar{\zeta}^{\bar{\jmath}}_{+} \zeta^{i}_{+} + \mtr{h}{a}{b} \bar{\eta}_{+}^{\bar{b}} \eta_{+}^{a}  +  \mtr{h}{i}{a} \bar{\eta}^{\bar{a}}_{+}  \zeta^{i}_{+} +  \mtr{h}{a}{\imath} \bar{\zeta}^{\bar{\imath}}_{+} \eta_{+}^{a} \right] \sum_{M}q_M \bar{\chi}_{-M}\chi_{-}^{M}
\Big\} ,
\end{align}
The covariant derivative in (\ref{eq:nlsm}) for chiral fields is defined as
\begin{align}
\nabla^{c}_{--} \psi_{+}^{p} &:= \pt_{--}\psi_{+}^{p} + \mcn{p}{q}{s}\, (\pt_{--} \phi^{q})  \psi^{s}_{+} 
\end{align}
where $\{\phi_{p}\}$ and $\{\psi_{p}\}$ are generic coordinates on the target space, say, $\{z_{i}, w_{a}\}$ and $\{\zeta_{+,i}, \eta_{+,a}\}$, respectively and $\mcn{p}{q}{s}$ is defined in Eq. (\ref{eq:mcnwpn}) while for fermi multiplet, it is shown that
\begin{align}
\nabla^{f}_{++} \chi_{-}^{M} &:= \pt_{++}\chi_{-}^{M} - i q_{M} \Omega_{++} \chi_{-}^{M}, 
\end{align}
with
\begin{align}\label{chc}
\Omega_{++} =\frac{i}{2H}~\left(\frac{1}{\varphi^2}z^{i}\olra{\pdvp}{\bar{z}_{i}}-\varphi^2 w^{a}\olra{\pdvp}{\bar{w}_{a}}\right).
\end{align}

Two remarks are in order here. First, we may wonder whether it is possible to enhance supersymmetry  in (\ref{eq:nlsm}) up to \ntwot  under an appropriate choice of parameters. The answer is negative. This can be traced back to the original construction of the gauged formulation, Eq. (\ref{eq:hetwcpn}). Evidently, the kinetic term of the left-handed fermions, $\tau_{+}$ and $\xi_{+}$ (corresponding to $\zeta_{+}$ and $\eta_{+}$), respectively, does not match that for the right-handed fermions $\chi_{-M}$. In addition, interactions of these fermions are  different. These two facts block the possibility of finding ${\cal N}=(2,2)$ models in the class of ${\cal N}=(0,2)$ models considered in this section. However, we will see that once the anomaly-free condition is met, the two-loop term in. the $\beta$ function vanishes much in the same way as
in ${\cal N}=(2,2)$, see Sec. \ref{seclo} for more details.

Second, in accordance with \cite{chens,ccsa}, we need to impose the constraints on the representation of the chiral and fermi multiplets for the theories to be free of the gauge anomalies, which implies their internal quantum consistency. Namely,
\begin{align}\label{af}
	\sum_{i,a} q^{2}_{i} + \tilde{q}^{2}_{a} = N + \tilde{N} = \sum_{M} q^{2}_{M}
\end{align}
where the $U(1)$ charges on the left-hand side come from the (left-handed) fermions in the supermultiplets $\mathcal{N}_{i}$ and $\varrho_{a}$ while those on the right-hand side are from the (right-handed) fermi multiplets. 

To wrap up this section, the geometry of the target manifold is identical to that obtained from the bosonic calculation, cf. Eq. (\ref{eq:wcpnmtr}) and (\ref{eq:mcnwpn}) --- (\ref{eq:scalar}), at the classical level. Since the fermion fields play no role in one-loop renormalization, the FI parameter and the K\"ahler potential receive the same corrections at the first order as discussed in the previous section. Making one step further, we will show in the next two sections that the \ntwoo case have no correction at the two-loop level.

\section{More on geometry of $\mathbb{WCP}(N,\tilde{N})$}
\label{app}
\setcounter{equation}{0}

\allowdisplaybreaks

As a complement to the discussion of $\mathbb{WCP}(N,\tilde{N})$ target manifold carried out above, here we will present the Riemann curvature tensors needed for the second loop to be obtained in Sec. \ref{seclo}.

For a generic K\"ahler manifold, the Riemann curvature tensor  can be written as 
\begin{align}
    \riec{p}{q}{r}{s} = -R_{q\overline{p}r}^{~ ~ ~ s} = \overline{\partial}_{\overline{p}}\mcn{s}{q}{r} \,,
\end{align}
implying
\begin{align}\label{rie}
    \riec{i}{j}{k}{l} &= 
    \frac{-1}{H \varphi^2}\left( \dlt{l}{k}\mtr{\delta}{j}{\imath} +\dlt{l}{j}\mtr{\delta}{k}{\imath}\right) +
    \frac{-r+2H}{H^3 \varphi^4}z_{\overline{\imath}}\left( \overline{z}_{j}\dlt{l}{k} +\overline{z}_{k}\dlt{l}{j} \right)\nonumber\\[2mm]
                      &+\frac{2}{H^2 \varphi^4}z^{l}\left( \overline{z}_{j}\mtr{\delta}{k}{\imath} +\overline{z}_{k}\mtr{\delta}{j}{\imath} \right)
                      -\frac{4(-r+2H)}{H^4 \varphi^6}z_{\overline{i}}\overline{z}_{j}\overline{z}_{k}z^l\,,\nonumber\\[2mm]
    \riec{i}{a}{j}{k} &= \frac{r}{H^{3}}z_{\overline{i}}\dlt{k}{j}\overline{w}_a -\frac{2}{H^2}\mtr{\delta}{j}{\imath}z^{k}\overline{w}_{a}+\frac{4(-r+H)}{H^4 \varphi^2}z_{\overline{\imath}}\overline{z}_{j}z^{k}\overline{w}_{a}\,,\nonumber\\[2mm] 
    \riec{i}{a}{b}{j} &= \frac{4r\varphi^{2}}{H^{4}}z^{j} z_{\overline{\imath}} \overline{w}_{a}\overline{w}_b\,,\nonumber\\[2mm] 
    \riec{i}{a}{b}{c} &= -\frac{r}{H^3}  z_{\overline{\imath}} \left(\bar{w}_{b}\dlt{c}{a}+\bar{w}_{a}\dlt{c}{b}\right)\,, \nonumber\\[2mm] 
    \riec{i}{b}{j}{a} &= \frac{\dlt{a}{b}}{H \varphi^{2}}\left[ \mtr{\delta}{j}{\imath}-\frac{-r+2H}{H^2 \varphi^{2}}\overline{z}_{j}z_{\overline{\imath}} \right] ,\nonumber\\[2mm] 
    \riec{i}{j}{k}{a} &= 0 \,,\nonumber\\[2mm] 
    \riec{a}{i}{j}{k} &= \frac{r}{H^3}\left(\bar{z}_{j}\dlt{k}{i}+\bar{z}_{i}\dlt{k}{j}\right)w_{\overline{a}}
    -\frac{4r}{H^4 \varphi^2}z^{k}\overline{z}_{j}\overline{z}_{i}w_{\overline{a}}\,,\nonumber\\[2mm] 
    \riec{a}{b}{i}{j} &= \frac{\varphi^2}{H}\left( \dlt{j}{i} - \frac{2}{H\varphi^2}z^{j}\overline{z}_i \right)\mtr{\delta}{b}{a}
-\frac{\varphi^4(2H+r)}{H^{3}}\dlt{j}{i}w_{\overline{a}}\overline{w}_{b}+\frac{4(H+r)\varphi^2}{H^4}\overline{z}_{i}z^{j}w_{\overline{a}}\overline{w}_b\,, \nonumber\\[2mm]
\riec{a}{b}{c}{i} &= \frac{2 \varphi^4}{H^2}z^{i} \left( \mtr{\delta}{b}{a}\overline{w}_c+\mtr{\delta}{c}{a}\overline{w}_b \right)
-\frac{4 \varphi^6(r+2H)}{H^4}z^{i}w_{\overline{a}}\overline{w}_b \overline{w}_c \,,\nonumber\\[2mm]
\riec{a}{b}{c}{d} &= -\frac{\varphi^2}{H}\left( \mtr{\delta}{b}{a}\dlt{d}{c}+\mtr{\delta}{c}{a}\dlt{d}{b}\right) 
+\frac{\varphi^4(2H+r)}{H^3}\left(\bar{w}_{b}\dlt{d}{c}+\bar{w}_{c}\dlt{d}{b}\right)w_{\overline{a}}\,,\nonumber\\[2mm]
    \riec{a}{b}{i}{c} &= -\frac{r}{H^3} \overline{z}_{i} w_{\overline{a}} \dlt{c}{b}\nonumber\\[2mm]
    \riec{a}{i}{j}{b} &= 0\,.
\end{align}
In what follows we will also need a special quadratic combination of the Riemann tensors,
\begin{align}\label{R2}
    R^{(2)}_{p\overline{q}} = R_{p~~t}^{~rs}\riec{q}{r}{s}{t} \,.
\end{align}

This combination can be obtained by a tedious although straightforward calculation.
Extensively employing (\ref{ivmtr}) and (\ref{rie}) we derive
\begin{align}\label{A5}
  R^{(2)}_{i\overline{\jmath}} &= \frac{2}{H^4\varphi^2}\left[ (N+\tilde{N}-1)H^2 +r^2 \right]\mtr{\delta}{i}{\jmath} + \frac{1}{H^7\varphi^4}\left[ -4(N+\tilde{N}-1)H^4 \right.\nonumber\\[2mm]
  &+4(N+\tilde{N}-1)H^3r -2(N+\tilde{N}+2)H^2r^2 +8Hr^3 -4r^4 \Big]\overline{z}_{i}z_{\overline{\jmath}}\,, \nonumber \\[2mm]
  R^{(2)}_{\imath\overline{a}} &= \frac{2r^2}{H^7}\left[ (N+\tilde{N}-2)H^2 + 2r^2 \right] \overline{z}_{i} w_{\overline{a}} \,,\nonumber\\[2mm]
  R^{(2)}_{a\overline{b}} &=  \frac{2\varphi^2}{H^4} \left[ (N+\tilde{N}-1)H^2+r^2 \right] \mtr{\delta}{a}{b} - \frac{\varphi^4}{H^7} \left[ 4(N+\tilde{N}-1)H^4 \right. \nonumber\\[2mm]
  &+2(2N+2\tilde{N}-3)H^3r + 2(N+\tilde{N})H^{2}r^{2}+6Hr^3+4r^4 \Big]\overline{w}_{a}w_{\overline{b}} \,.
\end{align}

One can easily verify the above expressions in two simple limiting cases. 
First, we consider the $\mathbb{CP}$ models and, then, the simplest example $N=\tilde{N}=1$ discussed in Sec. \ref{simples}. To reduce the generic case to 
$\mathbb{CP}(\tilde{N}-1)$, we should again take $N=0$ and $r$ negative, arriving at
\begin{align}
	\mtr{R^{(2)}}{a}{b} \to \frac{2\tilde{N}}{r^2}\mtr{g}{a}{b}.
\end{align}
On the other hand, addressing the $\mathbb{WCP}(1,1)$ model, we set $N=\tilde{N}=1$ and obtain
\begin{align}
	\mtr{R^{(2)}}{1}{1} \to \frac{4r^4}{H^7}\,,
\end{align}
cf. Eq. (\ref{612t}).

As is seen from Eq. (\ref{A5}), more and more structures emerge in higher order corrections. In comparison with the one-loop results in which only terms up to $H^{-4}$ show up at two loops we find additional $H^{-5}$ to $H^{-7}$ terms. It is not possible to absorb them in $\mtr{g}{p}{q}$. This illustrates our statement
of {\em non}-renormalizability of non-supersymmetric Hanani-Tong model. 

Similarly to the $\mathcal{N}=(2,2)$ case, the two-loop correction does not exist in \ntwoo sigma model since the imposition of (\ref{af}) will lead to a vanishing coefficient in front of of the second order term in the beta function.

\section{Second loop}
\label{seclo}
\setcounter{equation}{0}

Let us explore the renormalization of $\mathbb{WCP}(N,\tilde{N})$ in higher loops. For a given bosonic two-dimensional non-linear sigma model, the first two terms in the $\beta$ function are known in the general form (see e.g. \cite{SG}), namely,
\begin{align}\label{eq:beta}
	\mtr{\beta}{p}{q} =  \frac{1}{2\pi} \mtr{R}{p}{q}  
	+ \frac{1}{8\pi^2} \mtr{R^{(2)}}{p}{q} + \cdots
\end{align}
where the first term is nothing but the Ricci tensor and the following term represents the second power of the Riemann tensors, see Eq. (\ref{A5}).

Note that in Eq. (\ref{eq:beta}), the term proportional to the Ricci curvature stands for the one-loop correction while the second term
composed of  the square of the Riemann tensors relates to the two-loop calculation. The discussion of  the first order renormalization is presented in Sec. \ref{sec4}. Now we will briefly outline what happens in  the second order.

\begin{figure}[!htbp]
	\centering
	\includegraphics[width=.6\linewidth]{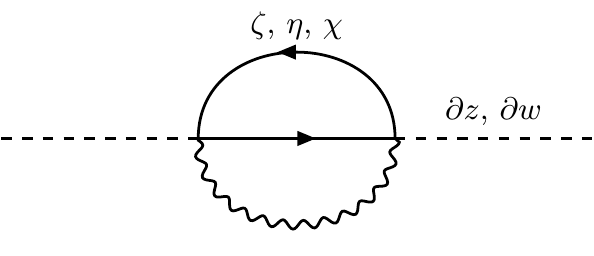}
	\caption{The second-order fermion loop diagram contributing to the $\beta$ function. The wavy line denotes the quantum part of the bosonic fields, $z$, and $w$.}
	\label{twolp}
\end{figure}
In the \ntwoo model we consider the  two-loop fermionic contribution shown in Fig. \ref{twolp}. If we work in the vicinity of the origin on the given patch
and keep only the lowest order terms, this is the only relevant diagram.
Then, it is easy to see that 
\begin{align}\label{tl}
-\frac{1}{16\pi^2} \left( N+\tilde{N} + \sum_{M} q^{2}_{M}\right)
=-\frac{1}{8\pi^2} \left( N+\tilde{N}\right).
\end{align}
In the above equality we employ the anomaly-free condition (\ref{af}), see \cite{chens} for details. It is important to stress that we should take the coefficient in front of the the fermion contribution to $\mtr{R^{(2)}}{p}{q}$ to be $-1/(16\pi^2)$ in the minimal ${\cal N}=(0,2)$ model \cite{CS}. The reason is that the  fermion graph in Fig. \ref{twolp} contributing at two loops acquires an extra factor $1/2$ in passing from the Driac to Weyl fermions. 

Returning to the anomaly-free non-minimal \ntwoo model we observe that the second order contributions from bosonic and fermionic fields cancel each other, and thus the second order coefficient vanishes. 

Indeed, if we neglect for a short while the $N$ and $\tilde{N}$ dependence in Eq. (\ref{tl}),  the fermionic two-loop correction reduces $-1/8\pi^{2}$. Together with the bosonic part in (\ref{eq:beta}), we obtain the second order coefficient 
\begin{align}
	-\frac{1}{8\pi^{2}} + \frac{1}{8\pi^{2}} = 0\, ,
\end{align}
leading to the same result as in  \ntwot model,  in which  only the first loop  survives as a result of a non-renormalization theorem.  The two models above are expected to have different contributions starting from three loops.

The remaining question refers to the overall factor  $N+\tilde{N}$  in Eq. (\ref{tl}). In the latter equation it was obtained by examining the vicinity of the origin of the given patch. 

Now we will have a closer look at the general form of  $\mtr{R^{(2)}}{p}{q}$ near the origin starting from (\ref{A5}). It is important that in the vicinity of the origin  of the origin $H \to r$ and therefore
\begin{align}\label{94}
R^{(2)}_{i\overline{\jmath}} &\to
(N+\tilde{N})\left[  \frac{2}{H^2\varphi^2}\mtr{\delta}{i}{\jmath} - \frac{1}{H^5\varphi^4} \left( 2r^2 \right)\bar{z}_{i} z_{\bar{\jmath}} \right]\,, \nonumber \\[2mm]
R^{(2)}_{\imath\overline{a}} &\to (N+\tilde{N})\left[ \, \frac{2r^2}{H^5} \overline{z}_{i} w_{\overline{a}} \right] \,,\nonumber\\[2mm]
R^{(2)}_{a\overline{b}} &\to 
(N+\tilde{N}) \left[ \frac{2\varphi^2}{H^2} \mtr{\delta}{a}{b} -
\frac{\varphi^4}{H^7}(10r^4)\,\overline{w}_{a}w_{\overline{b}}\right]\,,
\end{align}
Proportionality of $R^{(2)}_{p\overline{q}} $ 
 to the overall $N+\tilde{N}$ factor near the coordinate origin is obvious in the above expressions.
 
  Summarizing, from general covariance  and the above calculation in the non-minimal anomaly-free ${\cal N}=(0,2)$  $\mathbb{WCP}(N,\tilde{N})$ models we have 
\begin{align}\label{eq:beta2}
\beta\left(\mtr{g}{p}{q}\right) =  \frac{1}{2\pi} \mtr{R}{p}{q}  + \cdots \,,
\end{align}
where the ellipses stand for {\em three}-loop and higher order corrections.

In addition, we can compare with the results in \cite{chens} (see Eq. (4.10)). Generally speaking, the $\beta$ function in our notation  has the form
\begin{align}\label{gbeta}
	\beta(g^2) = -\frac{g^2}{4\pi} \left( \sum_{i} q_{i} - \frac{1}{2}\sum_{\alpha} q_{\alpha}\gamma_{\alpha} + \frac{1}{2} \sum_{M} q_{M}\gamma_{M}\right),
\end{align}
where $\gamma_{\alpha}$ and $\gamma_{M}$ are the anomalous dimensions of the chiral multiplets and the fermi multiplets, respectively. Also, the coupling constant $g$ is linked to the FI parameter through the relation
\beq
r=\frac{2}{g^{2}} \,.
\eeq
Perturbatively, to obtain the {\em two}-loop $\beta$ function, we only need $\gamma$ at the {\em one}-loop level, and we know that
\begin{align}
	 \gamma_{\alpha} \Big|_{\rm 1-loop} = 	 \gamma_{M} \Big|_{\rm 1-loop} \equiv \gamma \,.
\end{align}
Thus, Eq. (\ref{gbeta}) is simplified as
\begin{align}\label{gsb}
	\beta(g^2)_{\rm two-loop} = -\frac{g^2 \sum_{i} q_{i}}{4\pi}  + \frac{\gamma g^2}{8\pi}\left(\,\sum_{\alpha} q_{\alpha} -  \sum_{M} q_{M}\right).
\end{align}
Since this formula is universal, it is good enough to consider a simple example discussed in \cite{chens}, in particular, \ntwoo $\mathbb{CP}(N-1)$ model. In this case, there are $N$ chiral fields with positive unit charge and the same number of fermi multiplets with the same charge as that of chiral fields. As a consequence, the second term in (\ref{gsb}) vanishes and only the one-loop effect survives, namely,
\begin{align}
	\beta(g^2)_{\rm two-loop} = -\frac{N g^2}{4\pi}.
\end{align}

A similar argument can also be applied to the entire particular class of the ${\cal N}=(0,2)$ $\mathbb{WCP}(N,\tilde{N})$ models without internal anomalies which we consider in this paper. To proceed, let us first note that the anomaly-free condition (\ref{af}) in this model again forces the left-handed fermions to ``pair up" with the right-handed ones as is the case in \ntwot models. We can specify a particular choice for the set of $q_{M}$s such that $N$ of them have the $U(1)$ charge $+1$ and the rest $\tilde{N}$ fields have the  $U(1)$ charge  $-1$. Then, Eq. (\ref{gsb}) further reduces to
\begin{align}
		\beta\left(g^{2}\right)_{\rm two-loop} &= -\frac{(N - \tilde{N})g^{4}}{4 \pi}
		+ \frac{\gamma g^{2}}{8 \pi}\left[ N - \tilde{N} - (N - \tilde{N}) \right] \nonumber\\[2mm]
		&= -\frac{(N - \tilde{N})g^{4}}{4 \pi} \,,
\end{align}
i.e. the two-loop contribution vanishes in much the same way as we have seen in the $\mathbb{CP}(N-1)$ case. 
\vspace{3mm}

\section{Conclusions}
In this paper, we studied the structures of a particular NLSM derived from a class of GLSM and its $\mathcal{N} = (0,2)$ family. The geometry of such NLSM is a weighted complex projective space, $\mathbb{WCP}(N,\tilde{N})$, where $N$ and $\tilde{N}$ stand for the number of fields with the opposite $U(1)$ ``charges". This non-compact K\"ahlerian manifold does not admit a K\"ahler-Einstein metric which leads to emergence of extra structures and two different types of logarithms. 
Renormalization of the Fayet-Iliopoulos in  GLSM and that of the K\"ahler class in NLSM  coincide. However, there are additional logarithms in NLSM.

\vspace{5mm}

\section*{Acknowledgments}

The authors are grateful to Jin Chen, Sasha Gorsky, Sergey Ketov, Andrei Losev, David Tong and Alexey Yung for very very useful  discussions and correspondence.
This work  is supported in part by DOE grant DE-SC0011842.

\end{document}